\documentclass[]{aa} % for a referee version

\usepackage{graphicx}
\usepackage{natbib}
\usepackage{amsmath}
\usepackage{txfonts}
\usepackage{color}
\usepackage{enumerate}

\newcommand{\be}{\begin{eqnarray}}
\newcommand{\ee}{\end{eqnarray}}

\newcommand {\nbodypp}{\textsc{\mbox{nbody6\raise.4ex\hbox{\tiny++}}}}
\newcommand {\Msun} {\mbox{M$_{\odot}$}}

\begin{document}

\title{The expansion of massive young star clusters - \\observation meets theory}
\author{Susanne Pfalzner \& Thomas Kaczmarek}
\institute{
\inst{1}Max-Planck-Institut f\"ur Radioastronomie, Auf dem H\"ugel 69, 53121 Bonn, Germany\\
\email{spfalzner@mpifr.de}}
\date{ }

\titlerunning{Expansion of massive  young clusters}
\authorrunning{Pfalzner \& Kaczmarek}
\abstract
 {Most stars form as part of a star cluster. The most massive clusters in the Milky Way exist in two groups - loose and compact clusters - with significantly different sizes at the end of the star formation process. After their formation both types of clusters expand up to a factor 10-20 within the first 20 Myr.  Gas expulsion at the end of the star formation process is usually regarded as only possible process that can lead to such an expansion.}
{We investigate the effect of gas expulsion by a direct comparison between numerical models and observed clusters concentrating on clusters with masses  $>$10$^3$ \Msun. For these clusters the initial conditions before gas expulsion, the characteristic cluster development, its dependence on cluster mass, and the star formation efficiency
(SFE) are investigated . 
% as the entire dynamical expansion process is only observable for relatively massive clusters.
}
{We perform N-body simulations of the cluster expansion process after gas expulsion and compare the results with observations.  }
{We find that the expansion processes of the observed loose and compact massive clusters are driven by completely different physical processes. As expected the expansion of loose massive clusters is largely driven by the gas loss  due to the low SFE of $\sim$30\%. One new revelation is that all the observed massive clusters of this group seem to have a very similar size of 1-3 pc at the onset of expansion.  It is demonstrated that compact clusters have a much higher effective SFE of 60-70\% and are as a result much less affected by gas expulsion. Their expansion is mainly driven by stellar ejections caused by  interactions between the cluster members. The reason why ejections are so efficient in driving cluster expansion is that they occur dominantly from the cluster centre and over an extended period of time. Thus during the first 10 Myr the internal dynamics of loose and compact clusters differ fundamentally. }
 {}

\maketitle
\section{Introduction}

Star clusters\footnote{A brief comment on terminology: Some authors use the term "cluster" to refer only to stellar groups that remain bound after gas expulsion, while other authors use it for any significant stellar over-density regardless of its dynamical state. Here we use the word "cluster" in the latter sense.} form in the dense cores of giant molecular clouds. Such natal clusters consist of gas and stars with both components contributing to the total gravitational potential.  Usually  only some fraction of the gas will turn into stars, commonly referred to as star formation efficiency (SFE). Due to stellar feedback (e.g., Whitworth 1979, Pelupessy \& Portegies Zwart 2012, Dale et al. 2012), low-mass star outflows (Matzner \& McKee 2000), photo-ionisation of massive stars, and eventually explosions of the first supernovae (Zwicky 1953, Eggleton 2006), the gas may be rapidly removed from the young cluster. If the system is initially in virial equilibrium, the removal of the gas will leave the stars of the system in a supervirial state. This will cause some stars to become unbound and in case of a low SFE the entire cluster may dissolve. If a remnant cluster remains, it will expand in order to reach a new equilibrium state.

There exists a large body of work (for example, Tutukov 1978, Hills 1980, Lada, Margulis  \& Dearborn 1984, Adams 2000, Geyer  \& Burkert 2001, Kroupa, Aarseth  \& Hurley 2001, Boily  \& Kroupa 2003, Bastian  \& Goodwin 2006,  Baumgardt \& Kroupa 2007, L{\"u}ghausen et al. 2012)
investigating the conditions under which a bound remnant cluster remains after gas expulsion. 
It is found that the parameters that determine the outcome after gas expulsion are: the SFE,  the duration of the gas expulsion phase, and the distribution of gas and stars within the system. In case of the SFE, a critical value to form a remnant cluster is somewhere between 10\% and 35\%: the lower value corresponding to systems with slow gas expulsion  and/or sub-clustered structures; the higher value to instantaneous gas expulsion and smoothly distributed systems.
The current picture, suggested  by simulations, can be summarized as: a bound cluster is more likely to form if the SFE is high,  the time scale of gas removal long, and/or the system begins in a sub-virial state (e.g., McMillan et al. 2007, Allison et al. 2009, Goodwin 2009, Smith et al. 2011). For a more detailed review of existing theoretical and numerical work we refer to the accompanying paper (Pfalzner \& Kaczmarek 2013), in the following called Paper 1.

\begin{figure}[t]
\includegraphics[width=0.6\textwidth]{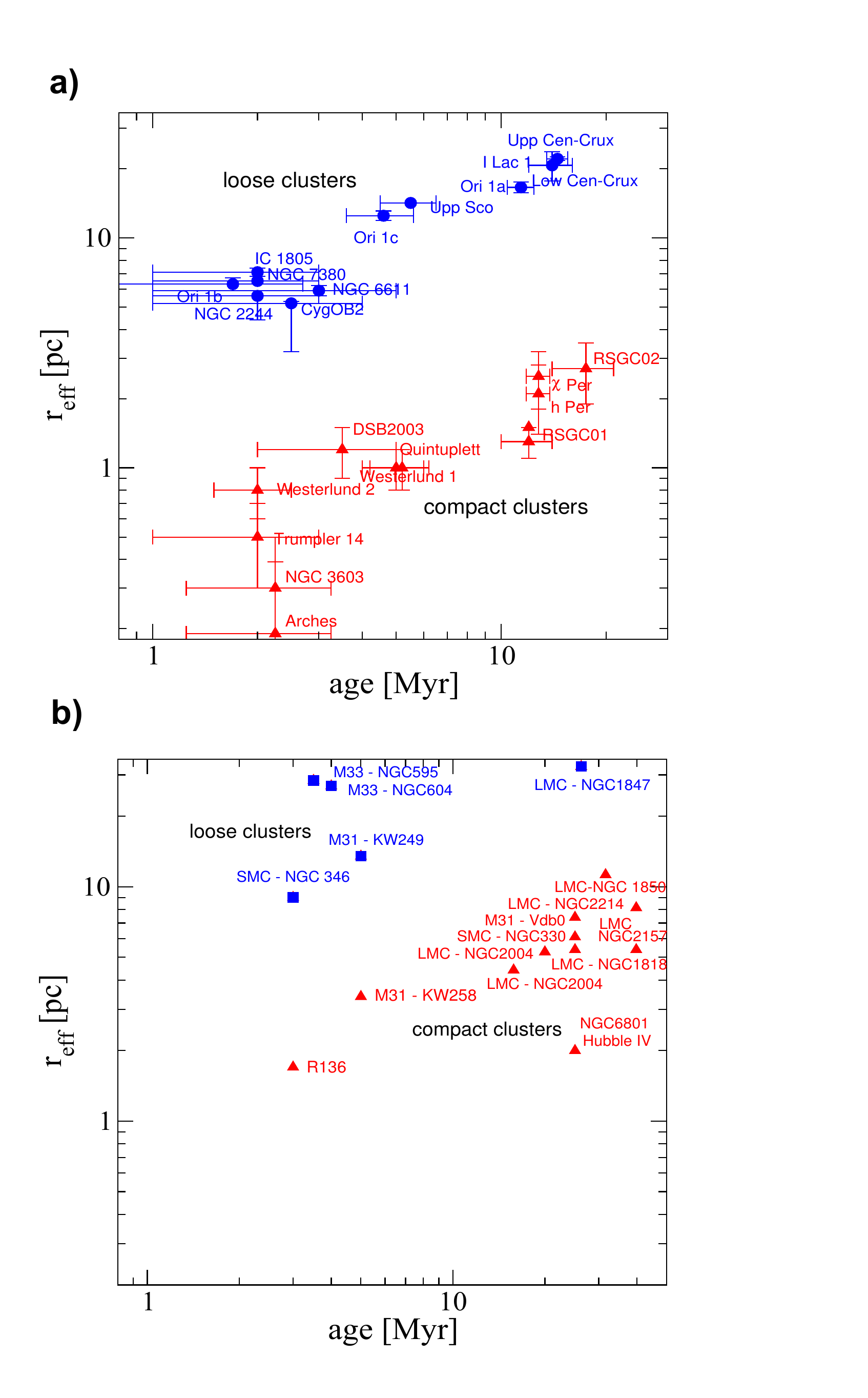}
\caption{Radii as a function of cluster age for a) massive Milky Way clusters (values from Pfalzner 2009) and extragalactic clusters in the Local Group (values from Portegies Zwart et al. 2010). }
\label{fig:milky}
\end{figure}

 There we demonstrated that in high-density clusters stellar ejections become so important that they lead to considerable additional mass loss which has to be taken into account when determining the fraction of bound mass after gas expulsion. This effect has been overlooked in the past as restricted computational resources were mostly compensated by either explicitly excluding encounters or implicitly by performing simulations where all particles representing the cluster stars had the same mass. Although some investigations of the gas expulsion phase included a full initial mass function (IMF) (Kroupa et al. 2001, Moeckel et al. 2012) they were usually limited to low-mass clusters simulating just a few specific cases. Here we use the same set of simulations as in Paper 1 but focus on the comparison with observations of the expansion phase of massive clusters ($>$ 10$^4$ \Msun ) after they have become largely devoid of gas.

From the observational side the effect of gas expulsion on young clusters is less clear.
In principle one should be able to reconstruct the details of the gas expulsion process from observations of clusters of different ages. However, for the low SFEs observed in the solar neighbourhood (SFE $<$ 35\%, Lada et al. 2010) a rapid increase in cluster size accompanied by a considerable loss of bound members can be expected. The result would be a dramatic decrease in surface density, so that low-mass clusters quickly drop below the detection limit. Therefore only clusters that were very massive  (${M_c} >$10$^3$ \Msun ) at the onset of gas expulsion, are still recognisable as such at later stages ($>$ 5 Myr). Thus we concentrate on such massive clusters which are largely devoid of gas.

Hunter (1999) and Maiz-Appelaniz (2001) pointed out that massive clusters seem to exist in two distinct groups - one more concentrated and the other more dispersed. As Fig.~\ref{fig:milky}  shows not only do two groups exist there seems to be two distinct correlations between cluster age and size. This could be interpreted as some massive clusters forming with large radii, remaining in that state and happening to be observed at older ages, which would be in accordance with the theories of, for example, Kruijssen et al. (2012). Alternatively, each type of cluster evolves along one of two fairly well-defined trajectories in the age-radius plane (Pfalzner 2009, Portegies Zwart et al. 2010, Gieles \& Portegies Zwart 2011) tracing the massive cluster expansion histories.  This is the interpretation adopted here.

It seems unlikely that the "gap" between the two groups of clusters (or "dearth" given the low number statistics) stems from observational biases as one finds exactly the same for clusters in the Local Group (Pfalzner \& Eckart 2009, Portegies Zwart et al. 2010). At even larger distances - outside the Local Group - the situation is less clear. The very different observational constraints in observations from inside the Galaxy vs. the outside view in the Local Group - speaks for the gap or dearth being real. This is what we assume in the following, when we try to put tighter constraints on the origin of the age-size relations.  Here we concentrate on the clusters in the Milky Way, whether the obtained results apply as well to extragalactic clusters will require further study. In Appendix B the sensitivity of the detection of the two cluster types on the used definitions of the cluster size is investigated.

The two cluster groups differ not only in the development of their size but as well in that of their masses. The compact clusters more or less retain their masses over the first 20 Myr of their development, whereas the loose clusters experience considerable mass loss (see Pfalzner 2011, Fig. 1). The combined effect of mass development and cluster expansion is reflected in an $R^{-3}$- dependence of the cluster density for the compact clusters in contrast to an $R^{-4}$-dependence for loose clusters. This has already been attributed to a much higher SFE in compact clusters (Pfalzner 2009). Recent observations of the velocity dispersions in compact clusters (Rochau et al.  2010, Mengel \& Tacconi-Garman 2007, Cottaar et al.  2012, Clarkson et al. 2012, Henault-Brunet et al. 2012) show that compact clusters have low-velocity dispersions despite their very young ages. This means that the clusters are likely close to their equilibrium state which indicates as well a high SFE (but see as well Banerjee \& Kroupa 2013 for a different approach).

\begin{table}
\begin{center}
\begin{tabular}{l*{5}{c}}
ID & $N_{stars}$ & $r_c^i$ & SFE &  $N(M)$ &$N_{sim}$ \\[0.5ex]
    &   & $[pc]$ &  & &\\[0.5ex]
\hline
LK  1     & 30 000      & 1.3  &  0.2-1.0    &  IMF        & 15\\
LK  1b     & 30 000      & 1.3  &  0.2-1.0  &  single    & 15\\
LK  2     & 30 000      & 4.6  &  0.2-0.3    &  IMF        & 15\\
LK  3     & 45 000      & 1.3  &   0.3          &  IMF        & 15\\
LK  4     & 15 000      & 1.3  &   0.3          &  IMF        & 31\\
LK  5     & 30 000      & 3.0  &   0.3          &  IMF        & 15\\
CK 1      & 30 000       & 0.1 &   0.3-0.7   & IMF        & 7\\
CK 2      & 30 000       & 0.2 &   0.6-0.7   & IMF        & 5\\

\end{tabular}
\caption{Properties of the presented cluster models: ID stands for the identifier, the second column depicts the number of stars, the third column denotes the initial cluster radius (for a definition see main Section 2), the fourth column shows the range of investigated SFEs,  $N(M)$ indicates the used mass representation and the last column the number of simulations performed for the given set-up. 
\label{table:sim}}

\end{center}
\end{table}

In the following we will compare the temporal development of various cluster properties expected from the simulations with the observed cluster properties. It will be shown that completely different processes dominate the cluster expansion in loose clusters like NGC 6611 compared to compact clusters like Westerlund 1. Namely, it is gas expulsion in the former case and ejection in the latter. It will be demonstrated why ejections are so efficient in driving cluster expansion in compact clusters.

\section{Method}

In the following we use the same set of simulations as in Paper 1, where the details of the model and the simulation can be found including a discussion of the approximations made. Here only a short summary is given:
Gas removal from massive clusters is considered to be effectively instantaneous, occuring in less than a crossing time (e.g. Goodwin 1997, Melioli \& de Gouveia dal Pino 2006). Therefore instead of explicitly simulating the gas expulsion process itself, we model the cluster as a stellar system out of virial equilibrium and follow the relaxation process (see Paper 1). A parameter $\epsilon$ is defined which describes how much  the cluster deviates from virial equilibrium after gas expulsion. If the cluster has been in virial equilibrium before gas expulsion, $\epsilon$ corresponds to  the SFE. 

In Paper 1 we show that it does not suffice to model such clusters with all stars having the same mass as often done in previous studies. Instead a full IMF has to be used, as otherwise the effect of encounters are severely under-represented.   We model the dynamics of the stars after the gas is expelled using the code Nbody6 (Aarseth 2003) which gives a high accuracy. In our simulations the stellar masses are chosen to represent the IMF as observed in young clusters (Kroupa 2001). 

The simulations were set up with all stars being initially single. This is a standard procedure which significantly reduces the computation time. However, during the simulations some binaries form usually with a preference for massive stars to catch a partner (Pfalzner \& Olczak 2007).

Our model stars are distributed so that the stellar density follows a $W_0 =$ 9 King profile. This is in accordance with observations which show that the stellar density profiles of young clusters just before gas expulsion are best represented by King models with $W_0>$ 7 (for example, Hillenbrand \& Hartmann 1998).

\begin{figure}[t]
\includegraphics[width=0.48\textwidth]{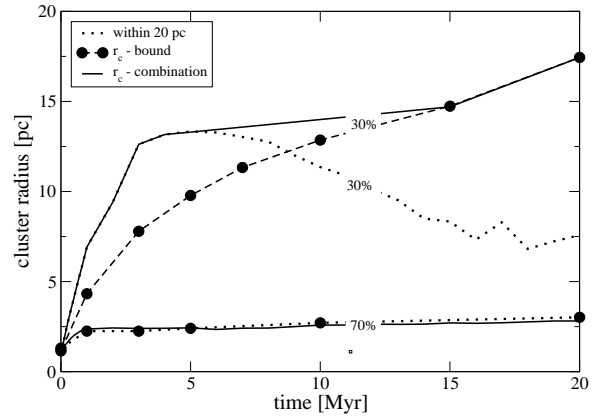}
\caption{Comparison of the temporal evolution of the radius of clusters with 30\% and 70\% SFE for model LK 1.
Two different methods of defining cluster membership are shown: i) from the content of bound and unbound stars within 20 pc (dotted line) and ii) from the total bound mass without spatial limitations (dashed line and circles).  The solid line shows the interpolation between i) and ii) which is used here to mimic the observations. }
\label{fig:hmr}
\end{figure}

To date it is still unclear whether the often
observed mass segregation in young clusters is primordial or  the result of rapid dynamical evolution. In addition, it is undecided whether only the most massive stars are concentrated at the cluster centre or if stars of all masses are affected by mass segregation (Andersen et al. 2011).  Given these uncertainties we treat the clusters here as initially non-mass segregated.  

The simulation campaign covered the parameter space from 10\% to 100\% SFE in steps of 10\%.  The parameters of the different cluster models are summarised in Table \ref{table:sim}. We performed simulations with different numbers of stars (15\,000, 30\,000, 45\,000)
and initial cluster radii (0.1 pc, 0.2 pc, 1.3 pc, 3 pc, 4.6\,pc). These values are motivated by typical observed properties of young ($<$ 4 Myr) massive clusters (Pfalzner 2009, Portegies Zwart et al. 2010, Gieles \& Portegies Zwart 2011) taking into account that for the used IMF the average stellar mass is 0.6 \Msun. This means, for example, that our simulations with 30\,000 particles correspond to a cluster mass of 18\,000 \Msun. Details of the parameter choice will be discussed in Sections 3. The simulations were performed for each parameter set repeatedly with random seeds for the initial distribution (see Table 1), so that the errors in the cluster radius are generally $<$ 3-4\%.

\section{Observed cluster sizes and masses}

A fundamental difficulty encountered comparing observed with simulated clusters in the gas expulsion phase is that the area occupied by bound and unbound stars overlaps at least for the first 5 to 10 Myr. Observationally it is difficult to distinguish between bound and unbound stars in this phase. In simulations usually only the development of the {\em bound} fraction is investigated. Figure \ref{fig:hmr} shows, as an example, the temporal development of the cluster radius of the bound fraction for our cluster model LK 1 for the cases of 30\% and 70\% SFE as dashed lines. Here the cluster radius is defined as the average distance of the B-type stars to the cluster centre. This definition corresponds to the method used in Wolff et al. (2007) for determining the radii of the observed clusters. Its relation to the half-mass radius and other size definitions is discussed in Appendix B. 

A comparison between this bound fraction and observations only makes sense for the initial conditions and at the end of the expansion process, but not during the gas expulsion process itself. One way to overcome this difficulty is to try to mimic the observational process in the simulations. Observations will initially identify all young stars  within a certain area as cluster members.  Such an aperture-limited approach can be mimicked in simulations by considering only stars within a predefined sphere.  Figure \ref{fig:hmr} shows the cluster radius obtained with this method within an aperture of 20 pc as dotted line.   
In case of 30\% SFE the aperture approach gives initially larger values for the radii than that of the bound stars. However, at $>$ 10\,Myr  aperture-limited value drops and eventually becomes smaller than obtained from the bound particles, simply because  a considerable amount of bound stars reside outside 20\,pc. 

In order to compare simulation results of the expansion process with observations neither of these two radius definitions are sufficient on their own. Instead we apply a combined method where we use the aperture approach until it reaches its maximum, the radius of the bound stars when it is larger than this value and in between we interpolate as indicated in Fig.~2 (solid line).  This takes care of the fact that, as soon as the unbound stars have moved far enough, no longer a confusion between bound and unbound members exists  for the observer. The cluster radius  thus obtained is henceforth denoted $r_c$. This method can only be regarded as a rough estimate to what is observed as there is no unique way how the cluster membership and radii are determined in the expansion phase in observations.

\subsection{Loose, extended clusters}

\begin{figure}[t]
\includegraphics[width=0.45\textwidth]{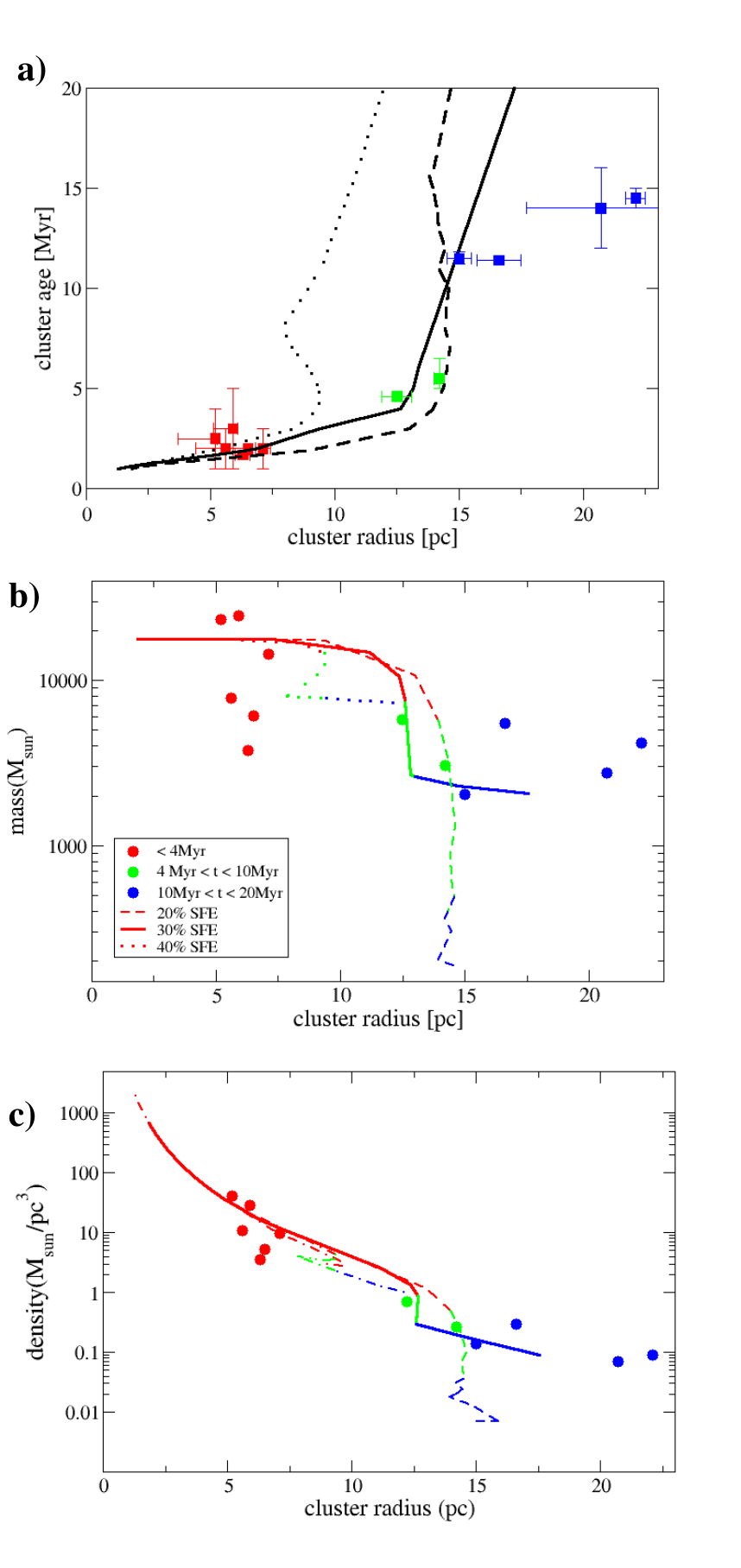}
\caption{Comparison between the simulations results for 20\% (dashed), 30\% (solid) and 40\% (dotted) SFE  for model LK 1 and the observed properties of young massive loose clusters (circles) as summarised in Pfalzner (2009).  a) shows the relation between cluster age and size $r_c$, b) the cluster mass and c) the cluster density as function of cluster size, respectively. The colours indicate cluster age (red $t_c<$ 4 Myr, green 4 Myr $< t_c < $ 10 Myr, blue $t_c > $10 Myr) where in the simulations it was assumed that the cluster had an age of 1\,Myr at the time of instantaneous gas expulsion.}
\label{fig:obs_sim_20_30_40}
\end{figure}

In the following we compare the simulation results to the masses and radii observed for loose clusters throughout the entire expansion process. Here we used the values from Wolff et al. (2007), as they are all determined the same way, so we have a {\em homogeneous} observational sample.

The simulation set LK 1 follows the temporal development of a cluster consisting of 30 000 stars with an initial  cluster radius of 1.3 pc at the moment of gas expulsion. Fig. \ref{fig:obs_sim_20_30_40} shows a comparison of the simulation results and the observed cluster properties of the loose cluster sequence. The simulation results for 20\%, 30\% and 40\% SFE are depicted as dashed, solid and dotted lines, respectively. 

The top panel a) shows the cluster radius as a function of cluster age. The cluster age was determined assuming that gas expulsion takes place instantaneously at an age of 1 Myr. In the initial phases ($<$ 3 Myr) all three curves (20\%, 30\% and 40\% SFE) of the age/size relation fit reasonably well. However, modelled clusters with 40\% SFE soon expand much slower than the observed cluster sequence, whereas the clusters with 20\% and 30\% have a fairly similar development in size with  the best agreement with the observed size-age development for 30\% SFE.
However, the oldest observed clusters are somewhat larger than all the simulated ones. 

The corresponding comparisons of the cluster mass and density ($\rho_c$ = 3$M_c$/4$\pi r_c^3$) as a function of cluster radius for different SFEs is shown in Fig. \ref{fig:obs_sim_20_30_40}b) and c), respectively. Here the comparison is in three parameters with the third one being the cluster age depicted by the colour scheme. Best fits between the simulations are again obtained  for 30\% SFE. 
If any loose clusters with SFEs of 40\% or higher formed,  the remaining cluster would be much smaller at an age of $>$ 10 Myr and would have a considerably higher density than observed in the sequence.  Such clusters have so far neither in the Milky Way nor in the Local Group been found, so that we deduce that a SFE$\sim$30\% is probably the upper limit for loose clusters. 

This finding is not completely unexpected as the maximum SFEs measured comparing stellar and gas masses in embedded clusters in the solar neighbourhood shows as well an upper limit of $\sim$ 30\% (Lada \& Lada 2003). However, they also find clusters with SFEs below 30\%. Looking at Fig. \ref{fig:obs_sim_20_30_40}c), the simulation results show that the density of clusters with 20\% SFE  declines rapidly after 4 Myr, so that these clusters would no longer be identified as overdensities at ages of $>$ {\mbox 5-10 Myr}. 

In our parameter study we varied as well the initial cluster mass and size (for details of the comparison see Appendix A).
We find that for the approximations adopted here, the best fit to
the cluster data would be obtained for an initial cluster radius of $r_c^i \sim$ 3 pc and 30\% SFE. However,  the neglected effects of binaries and stellar evolution likely lead to additional cluster expansion during these first 20 Myr of development with little additional mass loss. Taking this into account we conclude that the radii of the observed clusters were $r_c^i \sim$ 1-3 pc at the time of gas expulsion.

The comparison of simulated clusters with different initial mass basically shows that for initial cluster masses \mbox{$M_c^i<$ 10\,000 \Msun} the clusters  never reach a sufficient size that they could be the predecessors of the clusters observed at \mbox{10-20 Myr} (Fig.\,A.2). These clusters end up with radii well below l5 pc at an age of 20 Myr.

What fraction of clusters with $M_c^i>$10 000 \Msun\ does actually follow the observed cluster sequence? So far only a very limited number of clusters with sufficient information on masses, ages and sizes are known (for a list of possible Galactic massive cluster candidates see Appendix C). Nevertheless, our small sample might still be used to give a first estimate. Excluding the clusters with $M_c <$ 10 000 \Msun , it is only the 3 most massive clusters of the youngest age group that have the capacity to develop into the older observed population. This means in our crude sample statistics, that $>$ 50\% of all observed clusters with $M_c>$ 10 000 \Msun\ develop in a way that corresponds to simulations with 30\% $\pm$ 5\% SFE and an initial cluster radius in the range of 1-3 pc.  Thus for massive clusters it seems to be the rule rather than the exception to follow the observed loose cluster sequence.

\subsection{Compact clusters}

Now  we turn to the compact cluster sequence. In contrast to the homogenous sample of the loose cluster properties, the observational data of the compact cluster sequence come from different sources. This means that there exist slight differences in the way the cluster radii have been determined (see Appendix B). We continue with the same radius definition in our simulations as for the loose clusters. 

We take the observed mass and radius values at face value and model again clusters with 30\,000 stars but this time use a much smaller initial radius of $r_c^i$ = 0.1 pc (model CK 1). This initial radius corresponds to half the size of the smallest compact cluster (Arches) in our observational sample.  Here it is vital to explicitly include encounters by using a full IMF. Standard methods neglecting this lead to considerably less cluster expansion than expected from observations (see grey line Fig. 4a and details in Section 4). However, if one treats encounters correctly, good agreement between  observations and simulations with 60-70\% SFE (dashed and solid line in Fig.~4) are obtained. In these cases the clusters expand sufficiently while little mass is lost. 

In contrast to the loose clusters, 30\% SFE simulations (dotted line) this time do not match the observations at all as already after a few Myr the radius is far too large, reaching values \mbox{$\gg$ 10 pc} at the age of 10 Myr instead of the observed radii of \mbox{$\sim$ 1 pc}. Similarly, shortly after the gas expulsion ($2$ Myr)  the cluster mass  is an order of magnitude too small.  By contrast, clusters with larger SFEs ($\geq$80\%) do not expand sufficiently. 

\begin{figure}[t]
\includegraphics[width=0.45\textwidth]{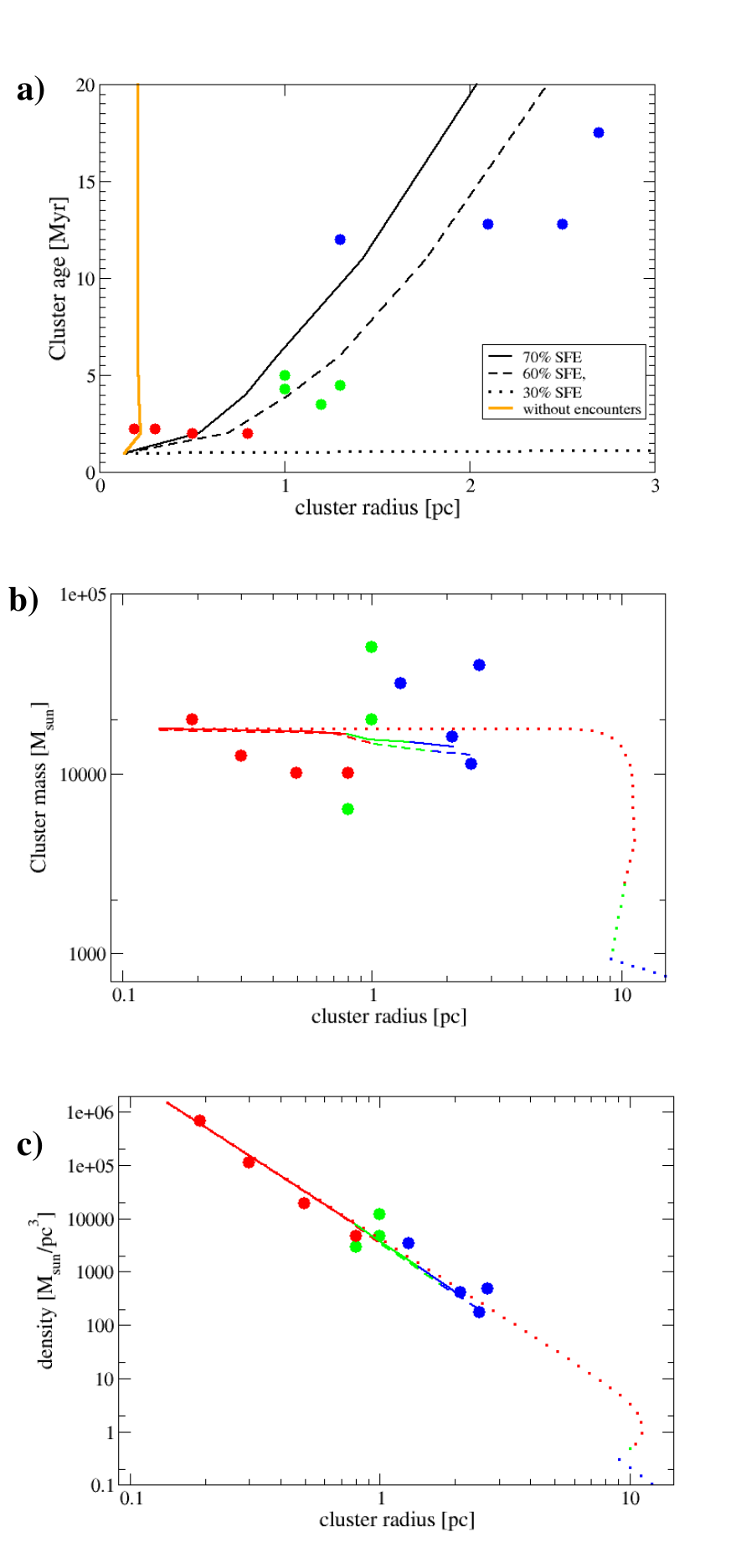}
\caption{Comparison between the simulations results for 30\% (dotteded), 60\% (dashed) and 70\% (solid) SFE for model CK 1 and the observed properties of young massive compact clusters (circles).  a) shows the relation between cluster age and size $r_c$, b) the cluster mass and c) the cluster density as function of cluster size, respectively. The colours indicate cluster age (red $t_c<$ 4 Myr, green 4 Myr $< t_c < $ 10 Myr, blue \mbox{$t_c >$ 10 Myr)} where in the simulations it was assumed that the cluster had an age of 1\,Myr at the time of instantaneous gas expulsion.}
\label{fig:starburst_SFE}
\end{figure}

The smallest cluster in our sample, the Arches cluster, with a cluster size of 0.1 pc, is located close to the Galactic Centre. Hence, it is exposed to a strong tidal field, which could in principle decrease the cluster size. Excluding Arches from our sample, we performed another set of simulations with the initial cluster radius twice as large as before (model CK 2). However, the results differ very little from before - again the 60-70\% SFE cases fit best and the cluster radius at 20 Myr is only marginally larger than in model \mbox{CK 1}.  

Here again, we can make a rough estimate of what proportion of massive compact clusters typically follows the sequential track. Comparing the number of clusters in the youngest age group with those in the oldest - 4 within 4 Myr compared to 4 within 10 Myr, leads to the conclusion that at least $\sim$ 40\% follow the compact cluster sequence. In other words, at least 40\% of compact clusters form with a star formation efficiency of 60-70\%.
Here again this has to be regarded as lower limit as cluster detection becomes more difficult in the 10-20 Myr age range than for younger clusters within the high density environments of the spiral arms and the Galactic Centre.

\section{Ejection-induced cluster expansion}

When modelling the early dynamics of compact clusters it is essential that encounters are included in the simulations of these very dense environments. Not only do the encounters lead to an additional mass loss  (see Paper 1) but, more importantly here, they are the dominant reason for the observed cluster expansion. We will detail that in the following.

\begin{figure}[t]
\includegraphics[width=0.5\textwidth]{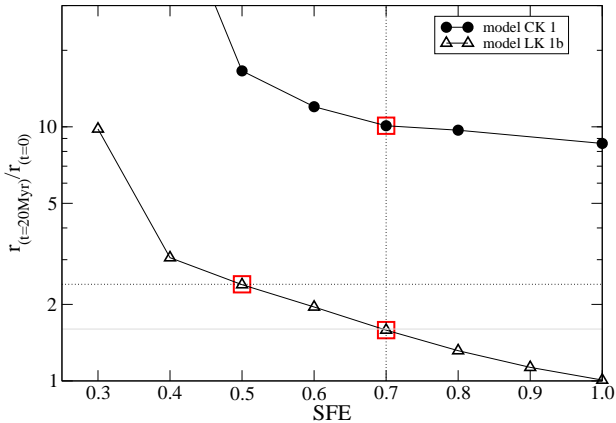}
\caption{ Comparison of the expansion factor as function of the SFE for model LK 1b (open triangles)
 and model CK 1 (filled circles). The squares highlight the values relevant for the comparison in section 4.}
\label{fig:expand1}
\end{figure}

Fig. \ref{fig:expand1} shows the factor by which our model cluster \mbox{CK 1}, which corresponds to a prototype compact cluster, has expanded after 20 Myr when simulated with (circles) and without (triangles) encounters. The latter is realised using model \mbox{LK 1b}, which would within the old picture just need rescaling.  The data points surrounded by squares in Fig.~5 highlight the here relevant values. First we look at the here relevant case of 70\% SFE - one would obtain a factor of 1.5-2 expansion without encounters compared to a factor of $\sim$ 10 if encounters are included. The latter matches the observational result for compact clusters. Thus in compact clusters it is actually the ejections that drive predominantly the cluster expansion, {\em not} the gas expulsion.

In Paper 1 it was demonstrated that in such dense clusters ejections account for an additional mass loss of $\sim$20\%. Thus one might conclude that this additional mass loss is responsible for the much stronger cluster expansion.  If the mass loss would just occur the same way as in the gas expulsion process, this would roughly correspond to a model with 50\% SFE. However,
Fig. \ref{fig:expand1} shows that also in case of 50\% SFE the cluster expands more (by approximately a factor 2.5), this is still far less than the observed factor of 10.

There are two reasons why ejections lead to a larger expansion than mass loss by gas expulsion:
i) ejections lead to mass loss from the cluster centre rather than from the outskirts of the cluster and ii) the mass loss by ejections is gradual rather than instantaneous. The former leads to a flattening of the potential at the cluster centre and the latter leaves the cluster enough time to adjust to the mass loss.  As a result of the two effects a relatively small mass loss can lead to considerable cluster expansion.

Stellar interactions also take place in loose clusters, but, due to their much lower stellar density, encounters are only responsible for $\sim$ 5\% of the total mass loss, whereas gas expulsion is responsible for 95\% of the mass loss (see Paper 1 for details). Thus the influence of ejections on loose clusters is limited - they are responsible for at most $\sim$25\% of the total cluster expansion. Moreover the cluster expansion of loose clusters is largely driven by the gas expulsion process.

Simplified, in loose clusters expansion is largely driven by gas expulsion, whereas in compact clusters largely ejections are responsible. This is the underlying physics that the cluster density can be approximated by a $R^{-4}$-dependence for loose clusters (Pfalzner 2009), whereas compact clusters seem to show a   $R^{-3}$-dependence. Ejections drive expansion despite little mass loss, whereas gas expulsion requires large mass loss to lead a large degree of expansion. The here presented results show that these scalings can only be regarded as rough estimates, the detailed simulations show a somewhat more complex behaviour.

\section{Discussion}

Here a short assessment of the assumptions made in above approach is required. The least critical is probably the neglect of stellar evolution.  Stellar evolution likely leads to some additional cluster expansion but predominantly in the later stages of development. It sets only in after a few Myr and should affect both cluster types to the same degree. So the general result should be relatively insensitive to stellar evolution.

The neglect of primordial binaries could have potentially a stronger effect. The binary dynamics in these early phases is complex (Parker et al. 2011, Marks et al. 2011, Kaczmarek et al. 2011, Korntreff et al. 2012). Including primordial binaries probably leads to more ejection events in both types of clusters. So ejections could become even more important. However, as this happens for both environments we expect that the general outcome - dominance of ejection for compact clusters vs. gas expulsion for loose clusters - will likely still hold. 

The influence of a possible primordial mass segregation is more difficult to anticipate. But again most likely it would lead to more ejections in both cases, but would not overturn the general result.

The high SFE of 60-70\% in compact clusters differs not only from that in loose clusters,  but as well from what one observes in embedded clusters in the solar neighbourhood. This means either  that close to the Galactic Centre and in the spiral arms much higher SFEs are common or that one sees only the central part of the young compact clusters where the SFE might be higher than in the unresolved outskirts of the cluster. Such a dependence of the SFE on the distance to the cluster centre
has been observed by Gutermuth et al. (2010), and was first suggested by Adams (2000). For a  recent application to massive clusters see Parmentier \& Pfalzner (2012).

Re-analysing the simulations by Bonnell et al (2008), Kruijssen et al. (2012) found that the simulated sub-clusters contained little gas and questioned whether gas expulsion is the reason behind cluster infant mortality.  They invoke tidal shocking as main cluster destruction mechanism. The low gas content, for example, in Westerlund 1 (Cottaar et al. 2012) could be interpreted as confirmation. However, there are two caveats with this: First, the investigated sub-clusters had an average mass of 40 \Msun,  that is nearly three orders of magnitude smaller masses than the clusters considered here. Whether a such massive scaling is realistic is questionable. Second,  there is no obvious process that would explain the rapid cluster expansion within the first 5 Myr of cluster development for the loose clusters other than interpreting it not as a sequence at all (see introduction). The two processes considered by Kruijssen et al. (2012), namely, stellar evolution and tidal perturbation usually lead to less expansion and on much longer time scales.

What happens to massive clusters after the first 20 Myr?  The observed, then called open, clusters are probably a mixture of remnants of compact as well as loose clusters. Both types of clusters reach their new nearly virial state already within 20 Myr. 

Afterwards, the loose clusters only expand slowly due to stellar evolution and binary interactions. As loose remnants are more extended and less massive than compact cluster remnants,  they are more susceptible to other disruptions like tidal fields. However, how strongly they are affected depends on their location in the Galaxy. Lada \& Lada (2003) report a loss of 96\% of clusters after 100 Myr, which means additional loss of $\sim$ 60\% of the clusters that survive until 10 Myr. During that time span probably preferentially the remnants of loose clusters are destroyed, as their large extend makes them susceptible to tidal fields.  Thus it could be expected that there is a trend of loose clusters only surviving for longer than 100 Myr in the outskirts of the Galaxy. 

For compact clusters the situation is different. At an age of 20 Myr ejections will still take place, although to a much lesser degree than before. Therefore cluster expansion due to ejections will gradually slow down. Due to their higher compactness, they will survive the effect of tidal fields much longer, with the exception of those close to the Galactic Centre.

\section{Conclusions}

The central aim of this paper was to investigate the underlying reason for the different  expansion behaviour of massive \mbox{($\geq$10$^4$ \Msun)} loose and compact clusters. Therefore we compared simulations of the cluster development after gas expulsion with observations throughout the first \mbox{20 Myr}. In short, although loose as well as compact clusters expand by about a factor of $\sim$10-20, two completely different processes are responsible - the expansion of loose clusters is caused by gas expulsion whereas that of compact clusters is largely driven by ejections.
In more detail that means:
For loose clusters, as perhaps to be expected, we find best fits in the development of the cluster size, mass and density over the first 20 Myr for models with a $\sim$ 30\% SFE. 
Current data indicate that all loose clusters seem to have a very similar size ($\sim$ 1-3 pc) at the onset of the expansion process. 
Due to the low SFE the gas expulsion process largely determines the following expansion.

More surprising are the findings for the compact clusters. The best fit between models and observations was found for  $\sim$ 60-70\% SFE and as such gas expulsion is much less important. Nevertheless these clusters do expand by a factor of  $\sim$10-20 within the first 20 Myr of their development. The reason is that stellar interactions between cluster members lead to ejections. Here cluster expansion is largely driven by encounter processes and gas expulsion is only of secondary importance. The ejection process is so efficient in driving cluster expansion, as in contrast to loose clusters cluster members are not predominantly lost from the cluster outskirts, but from the central cluster areas. In addition, ejections proceed over a longer time than the loss of gas ($\simeq$ 1 Myr). Therefore, a relatively small mass loss ($\sim$ 20\%)  leads to large cluster expansion.  

Open clusters in the age range 10-100 Myr constitute a mixture of remnant loose and compact clusters as despite loosing $\sim$90\% of their initial mass, loose clusters leave behind bound remnant clusters containing 1\,000-3\,000 stars. At ages $>$ 10 Myr they would be classified as open clusters with unusually large radii. 

The present work gives an explanation why compact clusters and loose clusters develop in different ways. However, what remains an unsolved question, is why massive clusters seem to form preferentially in two distinct sizes - as either compact clusters or more extended loose clusters. This seems to be the next challenge.

\acknowledgements

We wish to thank M. Messineo, G. Parmentier, and A. Stolte for very fruitful discussions. Part of the simulations were carried out at the J\"ulich Supercomputer Centre, Research Centre J\"ulich within Project HKU14.

\appendix

\section{Dependence on initial conditions}
In the following we demonstrate that the initial conditions chosen in model LK 1 represent the best fit to the observational data in the investigated parameter range.  

\subsection{Initial cluster size}

First we investigate the sensitivity of our results on the initial cluster radius. The mean cluster radius of the six youngest loose clusters is \mbox{6.1 pc}. As it is unlikely that the here presented massive exposed clusters are observed exactly at the end of the gas expulsion phase, we set up clusters with $\sim$1/4, 1/2, and 3/4 of this size and followed the cluster dynamics after gas expulsion.

\begin{figure}[t]
\includegraphics[width=0.45\textwidth]{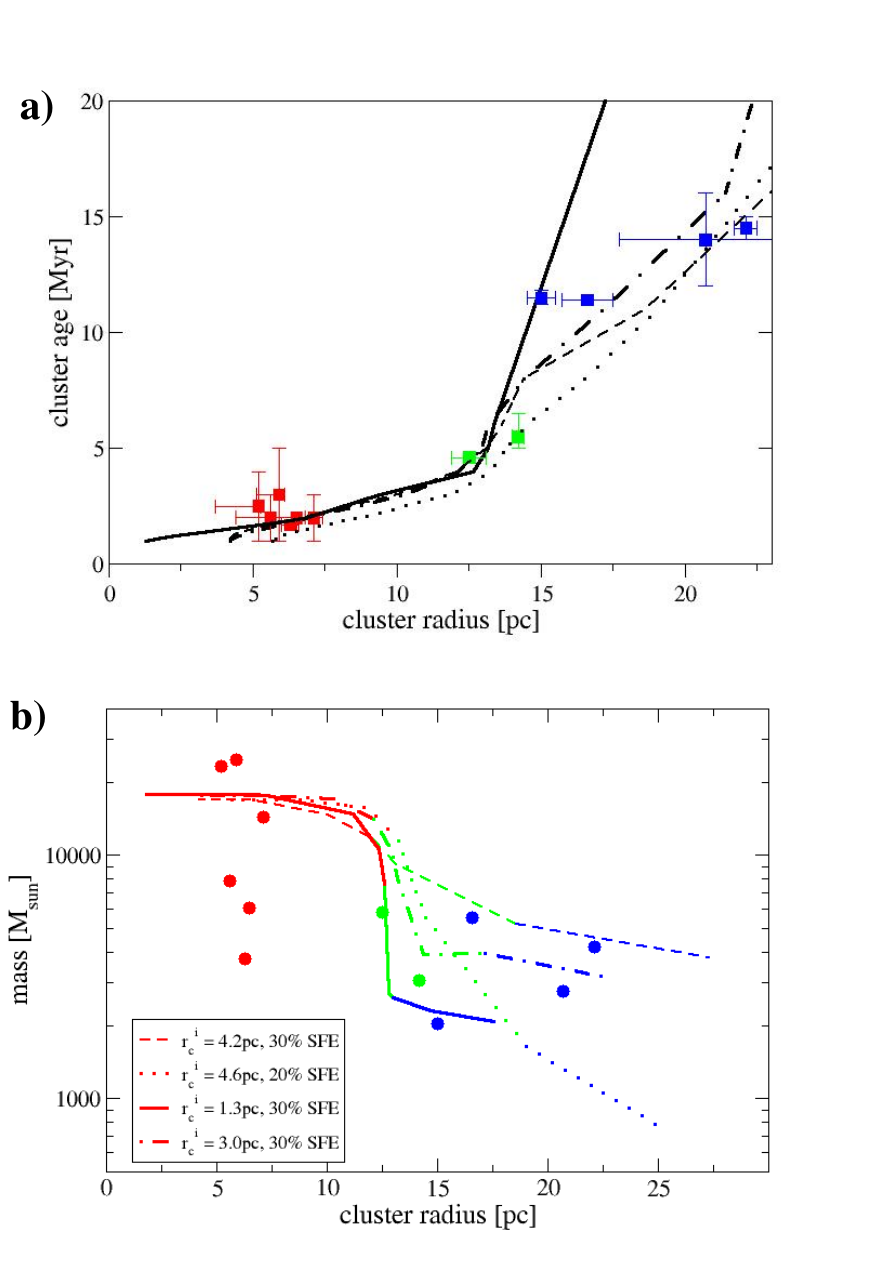}
\caption{Same as Fig. 3, but here for model LK1 only the simulation results for 30\% SFE are shown  (solid line). Those are compared to the results obtained using a larger initial radius (models LK 2 and LK 5 (dash-dotted line)). For model LK 2 the cases of 20\%  (dotted line) and 30\% (dashed line) SFE are depicted.}
\label{fig:comp4_6pc}
\end{figure}

In model LK 1 the initial radius was $r_c^i$ = 1.3 pc. In model LK 5 and LK 2 we use larger initial radii of \mbox{$r_c^i$ = 3 pc} and \mbox{$r_c^i$ = 4.6 pc}, respectively.  Fig. \ref{fig:comp4_6pc} shows that for $r_c^i$ = 4.6 pc and 30\% SFE the radius development fits the data quite well up to $\sim$ 10 Myr. In the later stages ($>$ 10 Myr) LK 1 (solid) matches only  the two smaller observed clusters, whereas LK 2 (dashed) fits better the two bigger ones. So from the size-age relation alone no clear distinction can be made which of the models gives the better fit.  
 
Looking at the mass development (Fig. \ref{fig:comp4_6pc}b) the first impression is that LK 2 would 
fit the results as well. However, closer inspection reveals that in the timespan 4-10 Myr the fit is relatively bad and, what is more severe, the curve partly fits the masses at the wrong cluster ages. Or in other words, the line fits but not the colour. The physical explanation is that larger initial cluster sizes lead to a slower cluster expansion resulting in a too high cluster mass at ages 4-10 Myr.
Similar problems are found for $r_c^i$ = 4.6 pc and 20\% SFE. Although the cluster mass
now fits better at 4-10Myr, at ages $>$ 10 Myr the obtained cluster masses are far too low, the mismatch in cluster age still remains and would not be remedied by using any other SFE either.

 The best fit to the cluster data is obtained for an initial cluster radius of $r_c^i \sim$ 3 pc (dash-dotted line). However,  the here neglected effects of binaries and stellar evolution likely lead to additional cluster expansion without much additional mass loss. Thus we conclude that the radius of these clusters likely was in the range 1-3 pc
at the end of the star formation process.

\subsection{Initial cluster mass}

\begin{figure}[t]
\includegraphics[width=0.45\textwidth]{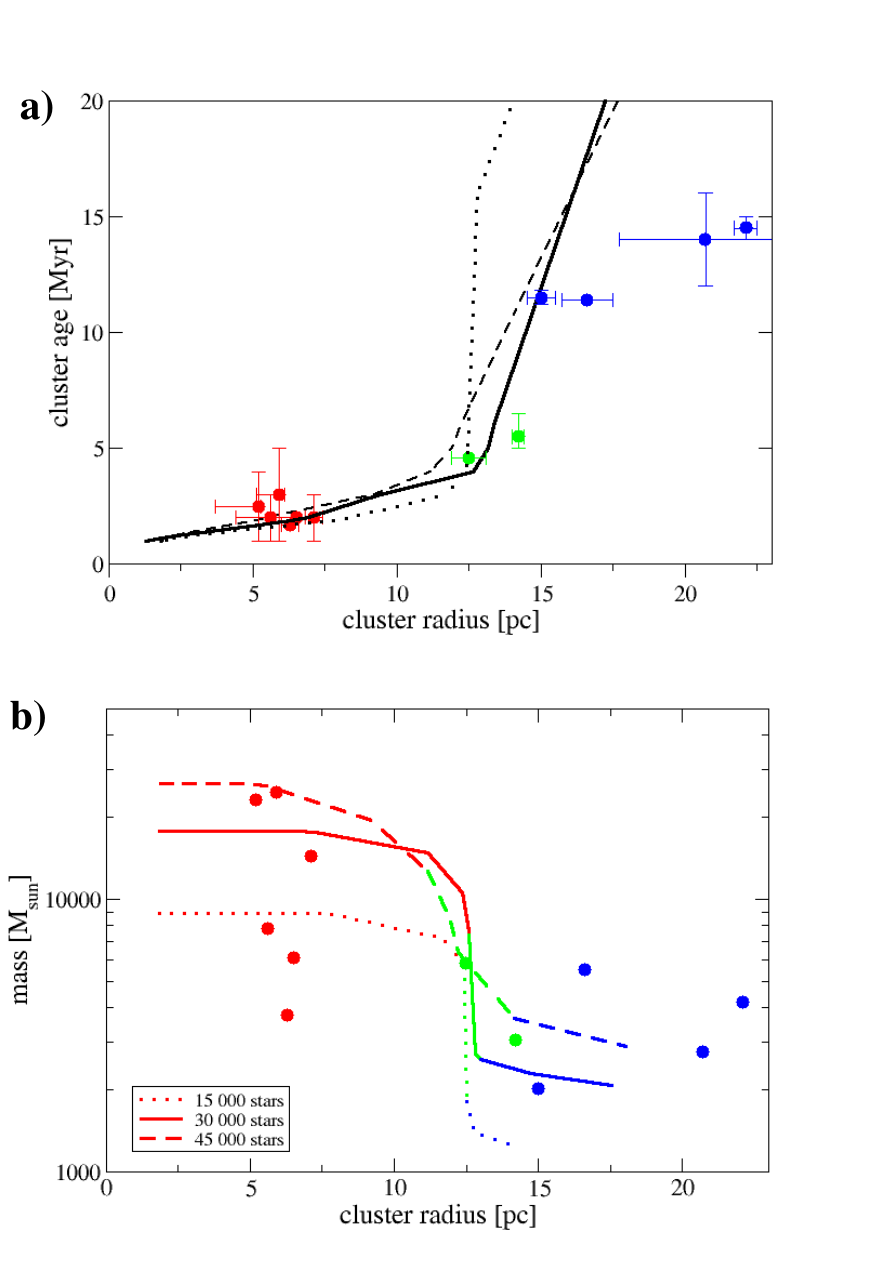}
\caption{Same as Fig. 3, but this time varying the initial cluster mass (model LK 1 - solid, model LK 3 - dashed, and model LK 4 - dotted. This corresponds to the models containing 30\,000, 45\,000 and 15\,000 stars, respectively. Only the simulation results for 30\% SFE are shown.}
\label{fig:comp_mass}
\end{figure}

So far we modelled clusters with 30\,000 stars which, for the IMF used here, is equivalent to a cluster mass $M_c \sim $ 18\,000 \Msun. This corresponds approximately to the {\em average} mass of the observed
young clusters in the sequence.  However,  clusters form with a wide variety of masses. Fig.  \ref{fig:comp_mass} shows the temporal development of clusters with initially 15\,000, 30\,000 and 45\,000 stars, corresponding to cluster masses of roughly 9\,000 \Msun, 18\,000 \Msun\, and 27\,000 \Msun. The comparison of the simulation results of these clusters of different initial mass shows that for initial cluster masses $M_c^i<$10\,000 \Msun\ the clusters  never reach a sufficient size that they could be the predecessors of the clusters observed at 10-20 Myr. These clusters end up with radii well below l5 pc and masses smaller than 2\,000 \Msun\ at an age of 20 Myr (see Fig.~A.2a).

This means clusters with $M_c^i <$ 10 000 \Msun\ do {\em  not} develop along the observed cluster sequence - at least for ages older than 10 Myr - , but follow their own sequence that is located at lower masses and radii. This has consequences for our statistical estimate in Section 3 - it is only the 3 most massive clusters of the youngest age group that have the capacity to develop into the older observed population. This means in our crude sample statistics, that $>$ 50\% of all observed clusters with $M_c>$ 10\,000 \Msun\ develop in a way that corresponds to simulations with 30\% $\pm$ 5\% SFE and an initial cluster radius in the range of \mbox{1-3 pc}.  So for such massive clusters it is the rule rather than the exception to follow the observed loose cluster sequence.  

\section{(In)sensitivity of the sequences to the radius definition}

\begin{figure}[t]
\includegraphics[width=0.52\textwidth]{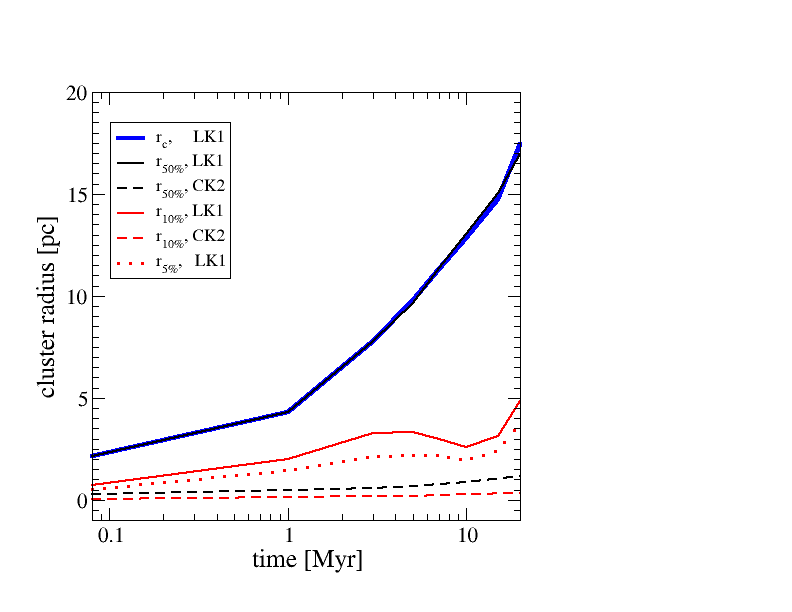}
\caption{Development of the radius of the bound cluster mass  
over time for the case of 30\% and 70\% SFE. The solid lines give the results for  clusters with $r_c^i =$ 0.1 pc and 70\% SFE (model CK 1) representative for compact clusters, and the dashed lines  indicate clusters with \mbox{$r_c^i$ = 1.3pc} (model LK 1) and 30\% SFE depicting loose clusters. The black lines indicate the half-mass radii, the red lines the 10\% Lagrange radii. The red dotted line indicates the 5\% Lagrange radius for the loose clusters. For the compact clusters the 5\% Lagrange radius would in the here chosen representation be basically undistinguishable from the 10\% Lagrange line.}
\label{fig:lagrange}
\end{figure}

As mentioned above, the loose cluster sequence is based on a homogeneous sample. The cluster sizes are all  determined in the same way as the average distance of the B-type stars to the cluster centre. More commonly used cluster definitions are the half-mass radius, the core radius and the radius where the cluster density equals the background density. Obviously the latter is unsuitable for a comparison of clusters in very different environments (solar neighbourhood, Galactic Centre, spiral arms) as done here. 

The core radius, i.e. the inner radius at which the surface density distribution flattens, is as well not really suitable to measure the cluster expansion in these early phases. The reason is twofold: i) the clusters are not in equilibrium and ii) the mixture of bound and unbound stars results in a cluster profile that is rather complex, so that the definition of a core can be an artefact of the current distribution of bound and unbound stars. For the compact clusters the situation is additionally complicated by the fact that the real core radius can be less than \mbox{0.1 pc}, which is for the more distant clusters very difficult to resolve.

Figure \ref{fig:lagrange} shows the development of the radii, which contain the innermost 5\% (red dotted line) and 10\% (red dashed and solid lines) of stars - 5\% and 10\% Lagrange radius - and the half-mass radius (black lines) for models LK\,1 (solid) and CK 1(dashed), being representative for loose and compact clusters. The core radius typically starts out close to the 10\% Lagrange radius and drops to approximately the 5\% Lagrange radius (Kroupa et al. 2001). The different expansion histories of compact (dashed) and loose (solid) clusters can be clearly distinguished for the half-mass radii (black), whereas this is more complex for the core radii (red): the core radius might even decrease for some time as seen in Fig. B1. Although the core radius at 20 Myr is larger than that at 1 Myr in both cases, the temporal development in between
is not so straight forward. For example, the core radius in model LK 1 at 10 Myr is smaller than at 5 Myr.
For the compact clusters the situation is even more complex: for a given half-mass radius the corresponding
core radii of the youngest clusters in the sample should be \mbox{$\ll$ 0.1 pc} which could in many cases be difficult to resolve.

\begin{table*}
\begin{center}
\begin{tabular}{l *{5}{c}}
ID                     & distance               &   age     &mass                     & source \\[0.5ex]
                        &  [kpc]                   &   [Myr]   & [10$^4$\Msun\ ]  &  [pc]    &             \\[0.5ex]
\hline
Masgomas 1     &   3.53$^{+1.55}_{-1.40}$  &          8-10  &        1.94$\pm$ 0.28                      & Ramirez Alegria et al. 2012\\
Stephenson 2   &   6                                      &                    &                                                   & Negueruela et al. 2012\\
Sh2-152          &   3.21 $\pm$ 0.21                 &                     &        0.245$\pm$ 0.079                 & Ramirez Alegria et al. 2011\\
C1 1813-178   &   4.8                 &         4-4.5  &        $\sim$ 1                           & Messineo et al. 2011\\
RSG3                &                                          &                     &        2-4                                 & Verheyen et al 2012\\
NGC 7419        & 2.3$\pm$ 0.3                       &     25$\pm$ 5  &                                        &  Verheyen et al 2012\\
GLIMPSE 13     &  1.6 - 4.3                           &                     &                                               &      Verheyen et al 2012\\
GLIMPSE 9       & 4.2                                    &     15-27      &         $>$ 0.16                        &     Verheyen et al 2012\\                              
SGR 1900+14 & 12.5 $\pm$ 1.7                     &   14              &                                           &Verheyen et al 2012\\
MFD2008        & 4                                       & 4.5               &             1                                & Messineo et al. 2008 \\
Trumpler 27    & 2.1$\pm$ 0.2                       & $<$ 6           &                                          & Verheyen et al 2012\\           
\end{tabular}
\caption{Properties of recently discovered massive clusters with $M_c >$ 10$^4$ \Msun. 
\label{table:clusters}}

\end{center}
\end{table*}

Thus, only the half-mass radius or the here used definition of the mean distance of the B stars are suitable to monitor the cluster expansion phase.
If there is no mass segregation in the cluster, both definitions lead to the same result. However, when the cluster expands the more massive stars remain bound more easily. As a result the half-mass radius is slightly smaller than the radius defined via the B stars at the end of the expansion process (see Fig. \ref{fig:lagrange}). However, the difference is less than 10\%. For initially mass segregated clusters the situation is probably more complex, which we plan to investigate in a future study.

A short remark to the data for the compact cluster. They come from different sources: The data for Arches, NGC 3603, Trumpler 14, Westerlung 1,2, RSGC1,2 and Quintuplett are from Figer (2008), the data for DBS2003 from Borissava et al. (2008), and the data for $\chi$ and h Per from Wolff et al. (2007). The latter define the radius as used above, whereas in Figer (2008) the radius is determined as the average projected separation from the centroid position. Given that most of these clusters are at distances where predominantly A or B stars are resolved, this means that very similar radius measures have been used in these two studies.  The only definition that differs considerably is that of Borissava et al. (2008), as there the radius was interpreted as the distance where the density distribution exceeds twice the standard deviation of the surface density in the surrounding.  So apart from the data point by Borissava et al. the used radius definitions are all very consistent for the sequence of compact clusters as well.

\section{Massive cluster candidates}

For the determination of the cluster expansion and mass loss after gas expulsion a large data base of the properties would be desirable. However, in the Milky Way currently the number of massive young compact and loose clusters with known masses and sizes is rather limited; basically restricted to the clusters displayed in Fig. 1. However, during the last few years a number of candidates of massive young clusters have been detected. Examples are given in Table B1. This list is far from complete, but shows that at least a doubling of the data points could be achieved in the near future. For these candidate massive clusters either the mass or cluster radius are currently unknown. This knowledge would help to answer the question, whether there exists a real size gap between loose and compact clusters or whether this region is just sparsely populated.  Whether a definite answer to this question is possible, depends on the total number of masive young clusters in the Milky Way. Currently it is estimated that this number is well below 100.

\end{document}